\documentclass[11pt]{article}
\pdfoutput=1
\usepackage[utf8]{inputenc}
\usepackage{braket,slashed,bm}
\usepackage{jheppub} \usepackage{braket,slashed,bm}
\usepackage{array,multirow}
\usepackage{amsmath}
\usepackage[normalem]{ulem}
\usepackage[T1]{fontenc}
\usepackage{makecell}
\usepackage{mathrsfs}
\usepackage{tikz}

\usepackage{subcaption}
\usetikzlibrary{arrows,decorations.pathmorphing,backgrounds,positioning,fit,petri,automata,shadows,calendar,mindmap,
decorations.markings,calc}
\usepackage{lscape}
\usepackage{hyperref}
\usepackage{graphicx}
\usepackage{booktabs}
\usepackage{floatrow}
\newfloatcommand{capbtabbox}{table}[][\FBwidth]
\usepackage{feynmp-auto}
\def\beq{\begin{equation}}
\def\eeq{\end{equation}}
\def\bea{\begin{eqnarray}}
\def\eea{\end{eqnarray}}

\def\hyp{\mathsf{y}}

\renewcommand{\d}{\delta}
\newcommand{\hst}{s_{\hat\theta}}
\newcommand{\hct}{c_{\hat\theta}}
\newcommand{\hsdt}{s_{2\hat\theta}}
\newcommand{\hcdt}{c_{2\hat\theta}}

\newcommand{\st}{s_\theta}

\def\hyp{\mathsf{y}}

\def\that{{\hat \theta}}

\def\bea{\begin{eqnarray}}
\def\eea{\end{eqnarray}}

\title{On interference and non-interference in the SMEFT}
\author{
Andreas Helset and Michael Trott\\
Niels Bohr International Academy and Discovery Centre, Niels Bohr Institute,
University of Copenhagen, Blegdamsvej 17, DK-2100 Copenhagen, Denmark}
\abstract{We discuss interference in the limit $\hat{m}_{W}^2/s \rightarrow 0$ in the Standard Model Effective Field Theory (SMEFT).
Dimension six operators that contribute to $\bar{\psi} \psi \rightarrow \bar{\psi'}_1 \psi_2' \bar{\psi'}_3 \psi'_4$ scattering events
can experience a suppression of interference effects with the Standard Model in this limit. This occurs for subsets of phase space
in some helicity configurations.
We show that approximating these scattering events
by $2\rightarrow 2$ on-shell scattering results for intermediate unstable
gauge bosons, and using the narrow width approximation, can miss interference terms present in the full phase space.
Such interference terms can be uncovered using off-shell calculations
as we explicitly show and calculate. We also study
the commutation relation between the SMEFT expansion and the narrow width approximation, and discuss some phenomenological implications of these
results.}
\begin{document}
\maketitle

\section{Introduction} \label{sec:intro}
When physics beyond the Standard Model (SM) is present at scales larger than the Electroweak scale,
the SM can be extended into an Effective Field Theory (EFT). This EFT can characterize
the low energy limit (also known as the infrared (IR) limit) of such physics relevant to the modification
of current experimental measurements. Assuming
that there are no light hidden states in the spectrum with appreciable couplings
in the SM, and that a $\rm SU_L(2)$ scalar doublet with hypercharge
$\hyp_h = 1/2$ is present in the IR limit of a new physics sector,
the theory that results from expanding in the Higgs vacuum expectation value
$\sqrt{2 \, \langle H^\dagger H} \rangle \equiv \bar{v}_T$ over the scale of new physics $\sim \Lambda$
is the Standard Model Effective Field Theory (SMEFT).

When the SMEFT is formulated using standard EFT
techniques,
this theoretical framework is a well defined and rigorous field theory
that can consistently describe and characterize the breakdown of the SM emerging from experimental measurements,
in the presence of a mass gap ($\bar{v}_T/\Lambda < 1$). For a review of such a
formulation of the SMEFT see Ref.~\cite{Brivio:2017vri}. The SMEFT
is as useful as it is powerful as it can be systematically improved, irrespective of its UV completion,
to ensure that its theoretical precision can match or exceed
the experimental accuracy of such measurements.

Calculating in the SMEFT to achieve this systematic improvement can be subtle. Well known subtleties in the SM predictions of cross sections can
be present, and further subtleties can be introduced due to the the presence of the
EFT expansion parameter $\bar{v}_T/\Lambda < 1$. Complications due to the
combination of these issues can also be present. As the SMEFT corrections to the SM
cross sections are expected to be small $\lesssim \%$ level perturbations, it is important to overcome these
issues with precise calculations, avoiding approximations or assumptions that introduce theoretical errors larger than the
effects being searched for, to avoid incorrect conclusions. For this reason, although somewhat counterintuitive,
rigour and precise analyses on a firm field theory footing are as essential in the SMEFT
as in the SM.

In this paper we demonstrate how subtleties of this form are present when calculating the leading interference effect
of some $\mathcal{L}^{(6)}$ operators as $\hat{m}_{W/Z}^2/s \rightarrow 0$.
We demonstrate how this limit can
be modified from a naive expectation formed through on-shell calculations due to off-shell
contributions to the cross section.
Furthermore, we show\footnote{For past discussions see
	Refs.~\cite{Berthier:2016tkq,Brivio:2017bnu,Brivio:2017vri}.} how to implement the narrow width approximation in a manner
consistent with the SMEFT expansion.

These subtleties are relevant
to recent studies of the interference of the leading SMEFT corrections in the $\hat{m}_{W/Z}^2/s \rightarrow 0$ limit, as they
lead to a different estimate of interference effects than has appeared in the literature when considering experimental observables.

\section{CC03 approximation of $\bar{\psi} \psi \rightarrow \bar{\psi'}_1 \psi_2' \bar{\psi'}_3 \psi'_4$} \label{sec:on-shell}
The Standard Model Effective Field Theory is constructed out of $\rm SU_C(3) \times SU_L(2) \times U_Y(1)$ invariant
higher dimensional operators built out of SM fields. The Lagrangian is given as
\bea
\mathcal{L}_{SMEFT} = \mathcal{L}_{SM} + \mathcal{L}^{(5)} + \mathcal{L}^{(6)} + \mathcal{L}^{(7)} + ...,
\quad \quad \mathcal{L}^{(d)}= \sum_{i = 1}^{n_d} \frac{C_i^{(d)}}{\Lambda^{d-4}} Q_i^{(d)} \hspace{0.25cm} \text{ for $d > 4$}.
\eea
We use the Warsaw basis \cite{Grzadkowski:2010es} for
the operators ($Q_i^{(6)}$) in
$\mathcal{L}^{(6)}$, that are the leading SMEFT corrections studied in this work.
We absorb factors of $1/\Lambda^2$ into the Wilson coefficients below.
We use the conventions of Refs.~\cite{Brivio:2017vri,Brivio:2017btx} for the SMEFT;
defining Lagrangian parameters in the canonically normalized theory
with a bar superscript, and Lagrangian parameters inferred from experimental measurements at tree level
with hat superscripts. These quantities differ (compared to the SM) due to the presence of higher
dimensional operators. We use the generic notation $\delta X = \bar{X} - \hat{X}$ for these
differences for a Lagrangian parameter $X$. See Refs.~\cite{Brivio:2017vri,Brivio:2017btx} and the Appendix
for more details on notation.

\begin{figure}[t]
	\centering
	\begin{subfigure}{0.20 \textwidth}
	\includegraphics[height=3.5cm,trim={ 0.5cm 0.0cm 0.0cm 0.0cm }, clip]{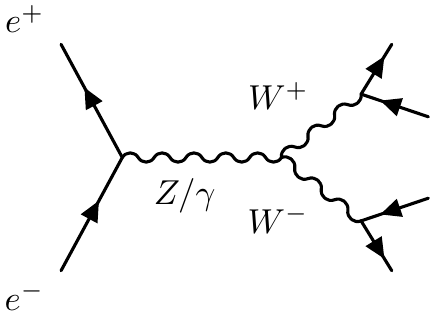}
	\label{fig:CC03_s_channel}
	\end{subfigure}
	\begin{subfigure}{0.40 \textwidth} \hspace{2cm}
	\includegraphics[height=3.5cm,trim={ 0.5cm 0cm 0cm 0cm }, clip]{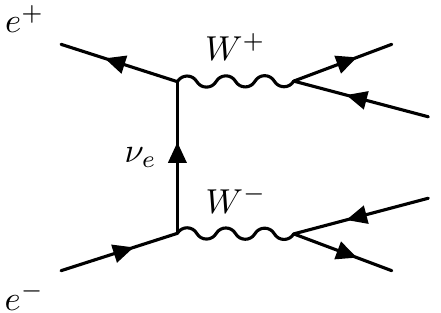}
	\label{fig:CC03_t_channel}
	\end{subfigure}

	\caption{The CC03 Feynman diagrams contributing to $\bar{\psi} \psi \rightarrow \bar{\psi'}_1 \psi_2' \bar{\psi'}_3 \psi'_4$ with leptonic initial states.}
	\label{fig:CC03}
\end{figure}

Consider $\bar{\psi} \psi \rightarrow \bar{\psi'}_1 \psi_2' \bar{\psi'}_3 \psi'_4$ scattering in the SMEFT with
leptonic $\bar{\psi} \psi$ and quark $\bar{\psi'}_1 \psi_2' \bar{\psi'}_3 \psi'_4$ fields.
The differential cross section for this process in the SM can be approximated by the
CC03 set of Feynman diagrams,\footnote{So named as CC indicates charged current.} where the $W^\pm$ bosons are considered
to be on-shell. This defines the related differential cross section $d \sigma(\bar{\psi} \psi \rightarrow W^+  W^-)/d \Omega$,
which is useful to define as an approximation to the observable, but it is formally unphysical as the $W^\pm$ bosons decay. The lowest order results of this form were
determined in Refs.~\cite{Tsai:1965hq,Flambaum:1974wp,Alles:1976qv,Bletzacker:1977gi,Gaemers:1978hg,Bilchak:1984ur,Hagiwara:1986vm,Beenakker:1994vn}
and the CC03 diagrams are shown in Fig.~\ref{fig:CC03}.
The amplitude for $\bar{\psi} \psi \rightarrow W^+ W^- \rightarrow \bar{\psi'}_1 \psi_2' \bar{\psi'}_3 \psi'_4$ in this approximation
is defined as
\bea
  \sum_{\substack{X= \{\nu,A,Z\} \\ \lambda^\pm = \{+,-\}}} \hspace{-0.5cm} M^{\lambda^\pm}_{X} = \bar D_W(s_{12})\bar D_W(s_{34})
  \mathcal{M}_{X}^{\lambda_i}
  \mathcal{M}_{W^+}^{\lambda_{12}}
  \mathcal{M}_{W^-}^{\lambda_{34}}, \quad  \bar D_W(s_{ij}) = \frac{1}{s_{ij} - \bar m_{W}^2 + i \bar \Gamma_W  \bar m_W + i\epsilon}, \nonumber \\
\eea
where a constant $s$-independent width for the $W^\pm$ propagators $\bar D_W(s_{ij})$ is introduced\footnote{We have checked and confirmed
that the novel interference effects we discuss below persist if an $s$ dependent width is used.} and
\begin{align*}
  \mathcal{M}_{\nu}^{\lambda_i} &=
  \mathcal{M}_{ee\rightarrow WW,\nu}^{\lambda_{12}\lambda_{34}\lambda_+\lambda_-} \delta_{\lambda_+}^{+} \delta_{\lambda_-}^{-}, \quad &
  \mathcal{M}_V^{\lambda_i} &= \mathcal{M}_{ee\rightarrow WW,V}^{\lambda_{12}
  \lambda_{34}\lambda_+\lambda_-}, \\
  \mathcal{M}_{W^+}^{\lambda_{12}} &= \mathcal{M}_{W^{+} \rightarrow f_1 \bar f_2}^{\lambda_{12}}, \quad &
  \mathcal{M}_{W^-}^{\lambda_{34}} &= \mathcal{M}_{W^- \rightarrow f_3 \bar f_4}^{\lambda_{34}},
\end{align*}
where $V = \{A,Z\}$. Here $\lambda_{12}$ and $\lambda_{34}$ label helicities of the intermediate $W^\pm$ bosons
with four momenta $s_{12},s_{34}$, and
$\lambda_{\pm}$ label helicities of
the $\bar{\psi} \psi$ initial state fermions. Transversely polarized massive vector bosons are labeled as $\lambda_{12/34} = \pm$
and  the remaining polarization (in the massless fermion limit) is labeled as $\lambda_{12/34} = 0$.
The individual sub-amplitudes are taken from Ref.~\cite{Berthier:2016tkq} where the complete SMEFT result
was reported (see also Refs.~\cite{Han:2004az,Corbett:2012dm,Corbett:2012ja,Corbett:2013pja,
Falkowski:2014tna,Falkowski:2016cxu,Falkowski:2015jaa,Butter:2016cvz,Englert:2014uua,Ellis:2014jta,Englert:2015hrx}).
The total spin averaged differential cross section is defined as
\begin{equation}
  \frac{d \, \sigma}{d\Omega \, ds_{12}ds_{34}} = \frac{\sum |M^{\lambda^\pm}_X|^2}{(2\pi)^2 \, 8s}, \quad \quad
  \sum |M^{\lambda^\pm}_X|^2 =|\bar D_W(s_{12})\bar D_W(s_{34})|^2
  \sum_{\substack{X= \{\nu,A,Z\} \\ \lambda^\pm = \{+,-\}}} M_X^{\lambda^\pm}
  (M_X^{\lambda^\pm})^*,
\end{equation}
where $d\Omega = d \cos \theta_{ab} \, d \phi_{ab} \, d \cos \theta_{cd} \, d \phi_{cd}  \, d \cos \theta \, d \phi$,
with $\theta,\phi$ the angles between the $W^+$ and $\ell^-$ in the center of mass frame.
The remaining angles describing the two body decays of the $W^\pm$ are in the rest frames of the respective bosons.
The integration ranges for $\{s_{12},s_{34}\}$ are $s_{34} \in \left[0, (\sqrt{s} - \sqrt{s_{12}})^2\right], s_{12} \in \left[0, s\right]$.
It is instructive to consider the decomposition of the general amplitude in terms of helicity labels of the initial state fermions,
and the intermediate $W^\pm$ bosons in the limit $\hat{m}_{W/Z}^2/s \rightarrow 0$ \cite{Gaemers:1978hg,Hagiwara:1986vm,Azatov:2016sqh,Falkowski:2016cxu,Azatov:2017kzw,Panico:2017frx}.
Note that the results we report below are easily mapped to other initial and finals states, so long as these states are  distinct.
\subsection{Near on-shell phase space}
First, consider the near on-shell region of phase space for the $W^\pm$ bosons defined by
\bea
{\rm \bf Case \, 1:} \hspace{1cm} s_{12} = s_1 \,\bar{m}_W^2,  \quad \quad s_{34} = s_3 \, \bar{m}_W^2.
\eea
This expansion is limited to the near on-shell region of phase space for the intermediate
$W^\pm$ bosons ($s_1 \sim s_3 \sim 1$) by construction.
Introducing $x = \hat{m}_W/\sqrt{s}$ and $y = s/\Lambda^2$  an expansion in
$x,y < 1$ can be performed by expressing the dimensionful parameters in terms of these dimensionless variables,
times the appropriate coupling constant when required.  The $\delta X$ parameters were rescaled to extract these dimensionful scales as
 $x^2 y \, \delta X = \bar{X} - \hat{X}$ where required.
This gives the results shown in Table \ref{Table:on-shell}.
\begin{table}[h!]
\centering
\tabcolsep 8pt
\renewcommand{\arraystretch}{2}
\begin{tabular}{|c|l|}
\hline
$\! \! \! \lambda_{12}\lambda_{34} \lambda_+\lambda_- \! \! \! $ & $\sum_X M^{\lambda^\pm}_{X}/4\pi\hat\alpha$  \\
\hline \hline
$00-+$ & $  \frac{\sin \theta}{2\sqrt{s_1s_3}}
			\left[\frac{1}{c_{\hat{\theta}}^2}
			+ \left(
  		\delta \kappa^{Z\alpha} -  \delta F_2^{Z\alpha} \right)y  \right]$
      \\
$\pm\pm-+$ & $  - \sin\theta \left[ \frac{x^2}{c_{\hat\theta}^2}  +\frac{y \,\delta \lambda^{Z\alpha}}{2}
+ \left( \delta g_{1}^{Z\alpha} -\delta F_2^{Z\alpha} - (s_1+s_3) \frac{\lambda_{Z\alpha}}{2}
				+ \frac{\delta \lambda_Z}{2 \, c_{\hat\theta}^2} \right) y \, x^2 \right]$ \\
$\pm 0-+$ & $ - \frac{(1\pm\cos\theta)\, x}{\sqrt{2s_3}} \left[\frac{1}{c_{\hat{\theta}}^2}
 + \frac{y}{2}
\left(\delta g_{1}^{Z\alpha} -2 \delta F_2^{Z\alpha} + \delta \kappa^{Z\alpha} + s_3 \,\delta \lambda^{Z\alpha} \right) \right]$\\
$0\pm-+$ &
$  \frac{(1\mp\cos\theta)\, x}{\sqrt{2s_1}} \left[\frac{1}{c_{\hat{\theta}}^2}
 + \frac{y}{2}
\left(\delta g_{1}^{Z\alpha} -2 \delta F_2^{Z\alpha}  + \delta \kappa^{Z\alpha} + s_1\delta \lambda^{Z\alpha} \right)
 \right]$
\\
$00+-$ & $  \frac{ \sin \theta}{2\sqrt{s_1s_3}} \left[(\frac{s^2_{\hat{\theta}} - c^2_{\hat{\theta}}}{2 \, c_{\hat{\theta}}^2 \,s_{\hat{\theta}}^2 })
+ \frac{s_1 + s_3}{2 \, s_{\hat{\theta}}^2}
+ \left(\delta \kappa^{Z\alpha}   - \frac{\delta \kappa_Z}{2 \, s_{\hat{\theta}}^2}
+ \frac{2 \, \delta \bar{g}_W^\ell}{s_{\hat{\theta}}^2} -\delta F_2^{Z\alpha}\right) y
\right]$\\
$\pm\pm+-$ & $
			- \frac{\sin\theta}{2}
			\left[\left(1-\frac{1}{2 s_{\hat\theta}^2}\right)
			\delta \lambda_{Z} - \delta \lambda_{\alpha} \right] y
		$ \\
$\pm 0+-$ & $
			\frac{(1\mp\cos\theta) \, x }{2 \sqrt{2s_3}}
      \left[(\frac{s^2_{\hat{\theta}} - c^2_{\hat{\theta}}}{c_{\hat{\theta}}^2 \,s_{\hat{\theta}}^2 })
				 + \frac{s_1}{s_{\hat\theta}^2}
				+\frac{s_3}{s_{\hat\theta}^2}
				\frac{1\pm2 + 3\cos\theta}{1+\cos\theta} - y  \frac{\left(\delta g_1^z  + \delta \kappa_z + s_3 \, \delta \lambda_z
                \right)}{2 s_{\hat\theta}^2}  \right.$
		 \\
	& $\left. \hspace{1.5cm}	- y \left(\delta F_1^{Z\alpha}  - \frac{4 \, \delta \bar{g}_W^\ell}{s_{\hat{\theta}}^2} - \left(
				\delta g_1^{Z\alpha} + \delta \kappa^{Z\alpha} + s_3 \, \delta \lambda^{Z\alpha} \right)\right)\right]$
\\
$0\pm +-$ & $
			- \frac{(1\pm\cos\theta) \, x }{2 \sqrt{2s_1}}
      \left[(\frac{s^2_{\hat{\theta}} - c^2_{\hat{\theta}}}{c_{\hat{\theta}}^2 \,s_{\hat{\theta}}^2 })
				+ \frac{s_3}{s_{\hat\theta}^2}
				+\frac{s_1}{s_{\hat\theta}^2}
				\frac{1\mp2 + 3\cos\theta}{1+\cos\theta}  - y \frac{\left(\delta g_1^z  + \delta \kappa_z + s_1 \, \delta \lambda_z
                \right)}{2 s_{\hat\theta}^2} \right.$
		 \\
	& $\left. \hspace{1.5cm}	- y \left(\delta F_1^{Z\alpha}  - \frac{4 \, \delta \bar{g}_W^\ell}{s_{\hat{\theta}}^2} - \left(
				\delta g_1^{Z\alpha} + \delta \kappa^{Z\alpha} + s_1 \, \delta \lambda^{Z\alpha} \right)\right)\right]$
\\
$\pm\mp+-$ & $
		\frac{(\mp1+ \cos\theta)\sin\theta}
		{2 \, s_{\hat\theta}^2\left(1+\cos\theta\right)}
		$ \\

\hline\end{tabular}
\caption{Expansion in $x,y < 1$ for the near on-shell region of phase space of the CC03 diagrams approximating
$\bar{\psi} \psi \rightarrow \bar{\psi'}_1 \psi_2' \bar{\psi'}_3 \psi'_4$. For exactly on-shell intermediate $W^\pm$ bosons
$s_1 = s_3 = 1$. We have used the notation $\delta F^i_{Z\alpha} = (\delta F_i^Z + \delta F_i^\alpha)/4\pi\hat\alpha$,
$\delta \lambda^{Z\alpha} = \delta \lambda_Z-\delta \lambda_\alpha$, $\delta \kappa^{Z\alpha} = \delta \kappa_Z-\delta \kappa_\alpha$
and  $\delta g_{1}^{Z\alpha} = \delta g_{1}^{Z} - \delta g_{1}^{\alpha}$.\label{Table:on-shell}}
\end{table}

Table \ref{Table:on-shell} shows an interesting pattern of suppressions
 to $\mathcal{L}^{(6)}$ operator corrections dependent upon the helicity configuration of the intermediate
 $W^\pm$ polarizations. This result is consistent with recent discussions
 in Refs.~\cite{Azatov:2016sqh,Falkowski:2016cxu,Azatov:2017kzw,Panico:2017frx}.
In the near on-shell region of phase space
a relative suppression of interference terms by $x^2$ for amplitudes with a $\pm$
polarized $W^\pm$ compared to the corresponding case with a $0$ polarization
is present.
These results for the $\lambda_{12}\lambda_{34} \lambda_+\lambda_- = \pm \pm + -$ and $\pm \pm - +$ helicity terms (which correspond
to initial state left and right handed leptons respectively) involve an
intricate cancellation of a leading SM contribution between the CC03 diagrams as
\bea
\frac{\mathcal{A}_{\pm\pm-+}}{4 \, \pi \, \hat{\alpha}} &\simeq&
-  \sin \theta \, \left[\left(1 + \delta \lambda_\alpha \, \frac{y}{2} \right)_{\alpha}
-  \left(1 + \delta \lambda_Z \, \frac{y}{2} \right)_{z} \right] + \cdots, \nonumber\\
&\simeq& - \frac{\sin \theta}{2} \left(\delta \lambda_\alpha - \delta \lambda_Z\right) \, y,  \\
\frac{\mathcal{A}_{\pm\pm+-}}{4 \, \pi \, \hat{\alpha}} &\simeq& - \, \sin \theta \, \left[\left(1 + \delta \lambda_\alpha \, \frac{y}{2} \right)_{\alpha \, pole}
-  \left(\left(1 - \frac{1}{2 \, s_{\hat{\theta}}^2} \right)\left(1 + \delta \lambda_Z \, \frac{y}{2} \right) \right)_{z \, pole}
- \left(\frac{1}{2 \, s_{\hat{\theta}}^2}\right)_{\nu} \right]  + \dots \nonumber \\
&\simeq& \frac{\sin\theta}{2}
\left[\left(1-\frac{1}{2 s_{\hat\theta}^2}\right)
\delta \lambda_{Z} - \delta \lambda_{\alpha} \right] y.
\eea
Here we have labeled the contributions by the internal states contributing to $M^{\lambda^\pm}_{X}$.
The $\{\nu, \alpha, Z \}$ contributions to the scattering events populate phase space in a different manner
in general. These differences are trivialized away in the near on-shell limit, leading to the cancellation
shown of the leading SM contributions in the expansion in $x$, but can be uncovered by considering
different limits of $s_{12},s_{34}$ and considering off-shell phase space.
\subsection{Both $W^\pm$ bosons off-shell phase space}
For example, consider the off-shell region of phase space defined through
\bea
{\rm \bf Case \, 2:} \hspace{1cm} s_{12} = s_1 \, s,  \quad \quad s_{34} = s_3 \, s,
\eea
with $s_1 \lesssim 1, s_3 \lesssim 1$. In this limit, one finds the expansions of the CC03 results
\bea
\mathcal{A}_{\pm\pm-+}^{s_1,s_3} &\simeq&
- 4 \, \pi \, \hat{\alpha} \, \sin \theta \, \sqrt{\tilde{\lambda}(s_1,s_3)} \left[\left(1 + \delta \lambda_\alpha \, \frac{y}{2} \right)_{\alpha}
-  \left(1 + \delta \lambda_Z \, \frac{y}{2} \right)_{z} \right] + \cdots,\\
\mathcal{A}_{\pm\pm+-} &\simeq& - 4 \, \pi \, \hat{\alpha} \, \sin \theta  \sqrt{\tilde{\lambda}(s_1,s_3)} \, \left[\left(1 + \delta \lambda_\alpha \, \frac{y}{2} \right)_{\alpha \, pole}
-  \left(\left(1 - \frac{1}{2 \, s_{\hat{\theta}}^2} \right)\left(1 + \delta \lambda_Z \, \frac{y}{2} \right) \right)_{z \, pole} \right], \nonumber \\
&+&\left[\left(\frac{4 \, \pi \, \hat{\alpha} \, \sin \theta}{2 \, s_{\hat{\theta}}^2 \, \sqrt{\tilde{\lambda}(s_1,s_3)}}\right)
\left(1 + \frac{-(s_1+s_3) + (s_1-s_3)(s_1-s_3 \mp \sqrt{\tilde{\lambda}(s_1,s_3)})}{1 - s_1 - s_3 + \sqrt{\tilde{\lambda}(s_1,s_3)} \cos \theta}\right) \right]_{\nu \, pole}.
\eea
Here we have defined $\sqrt{\tilde{\lambda}(s_1,s_3)} = \sqrt{1 - 2 \, s_1 - 2 s_3 - 2 s_1 \, s_3 + s_1^2 + s_3^2}$.
In the case of left handed electrons, the  differences in the way the various $t$ and $s$ channel poles populate phase space
are no longer trivialized away, and a SM contribution exists at leading order in the $x$ expansion. This
SM term can then interfere with the
contribution due to a $\mathcal{L}^{(6)}$ operator correction in the SMEFT.
The complete results in this limit for the helicity
eigenstates are reported in Table \ref{Table:bothoffshell}.

\begin{table}[h!]
\centering
\tabcolsep 8pt
\renewcommand{\arraystretch}{2}
\begin{tabular}{|c|l|}
\hline
$\lambda_i $& $\sum_X M^{\lambda^\pm}_{X}/4\pi\hat\alpha$ \\
\hline \hline
$00-+$ & $  \frac{\sqrt{\tilde{\lambda}} \, \sin \theta}{2\sqrt{s_1s_3}}
			\left[\frac{1}{c_{\hat{\theta}}^2}(1+ s_1 + s_3)
			+ \left(
  		\delta \kappa^{Z\alpha} - \delta F_2^{Z\alpha}(1+s_1 + s_3)
      +  \delta g_1^{Z\alpha}(s_1 + s_3) \right)y  \right] x^2$
      \\
$\pm\pm-+$ & $  - \sin\theta \sqrt{\tilde{\lambda}} \left[ \frac{x^2}{c_{\hat\theta}^2}  +\frac{y \,\delta \lambda^{Z\alpha}}{2}
+ \left(\delta g_{1}^{Z\alpha} -\delta F_2^{Z\alpha}
				+ \frac{\delta \lambda_Z}{2 \, c_{\hat\theta}^2} \right) y \, x^2 \right]$ \\
$\pm 0-+$ & $ - \frac{(1\pm\cos\theta)}{\sqrt{2s_3}} \left[\frac{x^2}{c_{\hat{\theta}}^2}   +\frac{y \, s_3  \,\delta \lambda^{Z\alpha}}{2}
 + \frac{y \, x^2}{2}
\left(\delta g_{1}^{Z\alpha} -2 \delta F_2^{Z\alpha} + \delta \kappa^{Z\alpha} + s_3 \,\delta \lambda^{Z\alpha} \right) \right]$\\
$0\pm-+$ &
$  \frac{(1\mp\cos\theta)}{\sqrt{2s_1}} \left[\frac{x^2}{c_{\hat{\theta}}^2}  +\frac{y \, s_1  \,\delta \lambda^{Z\alpha}}{2}
 + \frac{y \, x^2}{2}
\left(\delta g_{1}^{Z\alpha} -2 \delta F_2^{Z\alpha}  + \delta \kappa^{Z\alpha} + s_1\delta \lambda^{Z\alpha} \right)
 \right]$
\\
$00+-$ & $  - \frac{\sin \theta \, \sqrt{\tilde{\lambda}}}{4\sqrt{s_1s_3} \, s_{\hat{\theta}}^2} \left[
1+ s_1 + s_3 - \frac{1}{\tilde{\lambda}} \left(1 - (s_1- s_3)^2 - \frac{8 \, s_1 \, s_3}{1 - s_1 - s_3 +
\sqrt{\tilde{\lambda}} \cos \theta} \right) \right]$\\
$\pm\pm+-$ & $
\frac{\sin \theta \, \sqrt{\tilde{\lambda}}}{2 \, s_{\hat{\theta}}^2}
\left[1 - \frac{1}{\tilde{\lambda}}\left(1 + \frac{-(s_1+s_3)+ (s_1 - s_3)(s_1 - s_3 \mp\sqrt{\tilde{\lambda}} )}{1 - s_1 - s_3 + \sqrt{\tilde{\lambda}} \cos \theta} \right)
- s_{\hat{\theta}}^2 F_3(\lambda_{\alpha},\lambda_{Z}) y\right]$ \\
$\pm 0+-$ & $- \frac{(1\mp\cos\theta) \sqrt{\tilde{\lambda}}}{2 \sqrt{2s_3} \, s_{\hat{\theta}}^2}
\left[1 - \frac{1}{\tilde{\lambda}}\left(1 - s_1 + s_3 - \frac{2 \, s_3 (1 + s_1 - s_3 \mp
\sqrt{\tilde{\lambda}})}{1-s_1 - s_3 + \sqrt{\tilde{\lambda}} \cos \theta}\right)
- s_{\hat{\theta}}^2 s_3 F_3(\lambda_{\alpha},\lambda_{Z}) y \right]$
\\
$0\pm +-$ & $
			\frac{(1\pm\cos\theta) \sqrt{\tilde{\lambda}}}{2 \sqrt{2s_1} \, s_{\hat{\theta}}^2}
      \left[1 - \frac{1}{\tilde{\lambda}}\left(1  + s_1 - s_3 - \frac{2 \, s_1 (1 - s_1 + s_3 \pm
      \sqrt{\tilde{\lambda}})}{1-s_1 - s_3 + \sqrt{\tilde{\lambda}} \cos \theta}\right)
      - s_{\hat{\theta}}^2 s_1 F_3(\lambda_{\alpha},\lambda_{Z}) y \right]$
\\
$\pm\mp+-$ & $
		\frac{(\mp1+ \cos\theta)\sin\theta}
		{2 \, s_{\hat\theta}^2\left(1 - s_1 - s_3 +\sqrt{\tilde{\lambda}} \, \cos\theta\right)}
		$ \\

\hline\end{tabular}
\caption{Expansion in $x,y < 1$ for the off-shell region of phase space of the CC03 diagrams in
when $s_{12} = s_1 \, s, s_{34} = s_3 \, s$.
Here we have
used a short hand notation $\tilde{\lambda} = \tilde{\lambda}(s_1,s_3)$ and
$F_3(\lambda_{\alpha},\lambda_{Z}) = \left(\left(\frac{2 s_{\hat\theta}^2 -1}{2 s_{\hat\theta}^2}\right)
\delta \lambda_{Z} - \delta \lambda_{\alpha} \right)$ to condense results.
 \label{Table:bothoffshell}}
\end{table}

\subsection{One $W^\pm$ boson off-shell phase space}
One can define the region of phase space where one $W^\pm$ boson
is off-shell as
\begin{align*}
{\rm \bf Case \, 3a:} \hspace{1cm} s_{12} &= s_1 \, s, &   s_{34} &= s_3 \, \bar{m}_W^2, \\
{\rm \bf Case \, 3b:} \hspace{1cm} s_{12} &= s_1 \, \bar{m}_W^2, & s_{34} &= s_3 \, s,
\end{align*}
with $s_1 \lesssim 1, s_3 \sim 1$ for Case 3a, and $s_1 \sim 1, s_3 \lesssim 1$ for Case 3b. In these limits, the expansions of the CC03 results are
as follows. In Case 3a one has $\mathcal{A}_{\pm\pm-+}^{s_1,0}$  and
\bea
\mathcal{A}_{\pm\pm+-} &\simeq& - 4 \, \pi \, \hat{\alpha} \, \sin \theta  \sqrt{\tilde{\lambda}(s_1,0)} \, \left[\left(1 + \delta \lambda_\alpha \, \frac{y}{2} \right)_{\alpha \, pole}
-  \left(\left(1 - \frac{1}{2 \, s_{\hat{\theta}}^2} \right)\left(1 + \delta \lambda_Z \, \frac{y}{2} \right) \right)_{z \, pole} \right], \nonumber \\
&\,& \hspace{1cm} + \left[\left(\frac{4 \, \pi \, \hat{\alpha} \, \sin \theta \,}{2 \, s_{\hat{\theta}}^2 \, \sqrt{\tilde{\lambda}(s_1,0)}}\right)
\left(1 + s_1 - \frac{2 \, s_1(1- s_1 \pm \sqrt{\tilde{\lambda}(s_1,0)})}{1 - s_1 + \sqrt{\tilde{\lambda}(s_1,0)} \cos \theta}\right) \right]_{\nu \, pole}.
\eea
While in Case 3b one finds $\mathcal{A}_{\pm\pm-+}^{0,s_3}$ and
\bea
\mathcal{A}_{\pm\pm+-} &\simeq& - 4 \, \pi \, \hat{\alpha} \, \sin \theta  \sqrt{\tilde{\lambda}(0,s_3)} \, \left[\left(1 + \delta \lambda_\alpha \, \frac{y}{2} \right)_{\alpha \, pole}
-  \left(\left(1 - \frac{1}{2 \, s_{\hat{\theta}}^2} \right)\left(1 + \delta \lambda_Z \, \frac{y}{2} \right) \right)_{z \, pole} \right], \nonumber \\
&\,& \hspace{1cm} +  \left[\left(\frac{4 \, \pi \, \hat{\alpha} \, \sin \theta \,}{2 \, s_{\hat{\theta}}^2 \, \sqrt{\tilde{\lambda}(0,s_3)}}\right)
\left(1 + s_3 - \frac{2 \, s_3(1- s_3 \mp \sqrt{\tilde{\lambda}(0,s_3)})}{1 - s_3 + \sqrt{\tilde{\lambda}(0,s_3)} \cos \theta}\right) \right]_{\nu \, pole}.
\eea
Again, the SM term for left handed initial states does not vanish and can interfere with the
contribution due to a $\mathcal{L}^{(6)}$ operator correction in the SMEFT in these regions of phase space.
The complete results in this limit for the helicity eigenstates are reported in Table \ref{Table:partialonoffshella},\ref{Table:partialonoffshellb}.
\begin{table}[h!]
\centering
\tabcolsep 8pt
\renewcommand{\arraystretch}{2}
\begin{tabular}{|c|l|}
\hline
$\! \! \! \lambda_i \! \! \! $ & $\sum_X M^{\lambda^\pm}_{X}/4\pi\hat\alpha$ \\
\hline \hline
$00-+$ & $  \frac{\sqrt{\tilde{\lambda}} \, \sin \theta}{2\sqrt{s_1s_3}}
			\left[\frac{1}{c_{\hat{\theta}}^2}(1+ s_1)
			+ \left(
  		\delta \kappa^{Z\alpha} -  \delta F_2^{Z\alpha}(1+s_1)
      +  \delta g_1^{Z\alpha} s_1 \right)y  \right] x$
      \\
$\pm\pm-+$ & $ - \sin\theta \sqrt{\tilde{\lambda}} \left[ \frac{x^2}{c_{\hat\theta}^2}  +\frac{y \,\delta \lambda^{Z\alpha}}{2}
+ \left(\delta g_{1}^{Z\alpha} -\delta F_2^{Z\alpha}
				+ \frac{\delta \lambda_Z}{2 \, c_{\hat\theta}^2}  - \frac{\delta \lambda^{Z\alpha} (1 + s_1) s_3}{2 \, \tilde{\lambda}} \right) y \, x^2 \right]$ \\
$\pm 0-+$ & $- \frac{(1\pm\cos\theta)\, \sqrt{\tilde{\lambda}} \, x}{\sqrt{2s_3}} \left[\frac{1}{c_{\hat{\theta}}^2}
 + \frac{y}{2}
\left(\delta g_{1}^{Z\alpha} -2 \delta F_2^{Z\alpha} + \delta \kappa^{Z\alpha} + s_3 \,\delta \lambda^{Z\alpha} \right) \right]$\\
$0\pm-+$ &
$  \frac{(1\mp\cos\theta) \sqrt{\tilde{\lambda}}}{\sqrt{2s_1}} \left[\frac{x^2}{c_{\hat{\theta}}^2}  +\frac{y \, s_1  \,\delta \lambda^{Z\alpha}}{2}
 + \frac{y \, x^2}{2}
\left(\delta g_{1}^{Z\alpha} -2 \delta F_2^{Z\alpha}  + \delta \kappa^{Z\alpha} + \frac{s_1 \delta \lambda_Z}{c_{\hat\theta}^2}
- \frac{s_1 \, s_3 \, (1+ s_1)}{(1- s_1)^2}\,\delta \lambda^{Z\alpha}  \right)
 \right]$
\\
$00+-$ & $ - \frac{\sin \theta}{\sqrt{s_1 \, s_3} \,  \sqrt{\tilde{\lambda}} \, s_{\hat\theta}^2} \, \frac{s_1 (s_1^2-1)}{4 \, x}$\\
$\pm\pm+-$ & $
\frac{ \sin \theta \, \sqrt{\tilde{\lambda}}}{2 \, s_{\hat{\theta}}^2}
\left[1 - \frac{1}{\tilde{\lambda}}\left(1   - \frac{s_1 (1 - s_1  \pm s_1
\sqrt{\tilde{\lambda}})}{1-s_1 + \sqrt{\tilde{\lambda}} \cos \theta}\right)
- s_{\hat{\theta}}^2  F_3(\lambda_{\alpha},\lambda_{Z}) y\right]$ \\
$\pm 0+-$ & $- \frac{s_1 (s_1 - 1)(1 \mp \cos \theta)}{2 \, \sqrt{2 \, s_3} \, s_{\hat{\theta}}^2 \,  \sqrt{\tilde{\lambda}}} \, \frac{1}{x}$
\\
$0\pm +-$ & $
			\frac{(1\pm\cos\theta) \sqrt{\tilde{\lambda}}}{2 \sqrt{2s_1} \, s_{\hat{\theta}}^2}
      \left[1 - \frac{1}{\tilde{\lambda}}\left(1  + s_1  - \frac{2 \, s_1 (1 - s_1  \pm
      \sqrt{\tilde{\lambda}})}{1-s_1 + \sqrt{\tilde{\lambda}} \cos \theta}\right)
      - s_{\hat{\theta}}^2 s_1  F_3(\lambda_{\alpha},\lambda_{Z}) y \right]$
\\
$\pm\mp+-$ & $
		\frac{(\mp1+ \cos\theta)\sin\theta}
		{2 \, s_{\hat\theta}^2\left(1 - s_1 +\sqrt{\tilde{\lambda}} \, \cos\theta\right)}
		$ \\

\hline\end{tabular}
\caption{Expansion in $x,y < 1$ for the off-shell region of phase space of the CC03 diagrams.
Here  we have
used a short hand notation $\tilde{\lambda} = \tilde{\lambda}(s_1,0)$. \label{Table:partialonoffshella}}
\end{table}
\begin{table}[h!]
\centering
\tabcolsep 8pt
\renewcommand{\arraystretch}{2}
\begin{tabular}{|c|l|}
\hline
$\! \! \! \lambda_i \! \! \! $ & $\sum_X M^{\lambda^\pm}_{X}/4\pi\hat\alpha$ \\
\hline \hline
$00-+$ & $  \frac{- \sqrt{\tilde{\lambda}} \, \sin \theta}{2\sqrt{s_1s_3}}
			\left[\frac{1}{c_{\hat{\theta}}^2}(1+ s_3)
			- \left(
  		\delta \kappa^{Z\alpha} +  \delta F_2^{Z\alpha}(1+s_3)
      +  \delta g_1^{Z\alpha} s_3 \right)y  \right] x$
      \\
$\pm\pm-+$ & $ - \sin\theta \sqrt{\tilde{\lambda}} \left[ \frac{x^2}{c_{\hat\theta}^2}  +\frac{y \,\delta \lambda^{Z\alpha}}{2}
+ \left(\delta g_{1}^{Z\alpha} -\delta F_2^{Z\alpha}
				+ \frac{\delta \lambda_Z}{2 \, c_{\hat\theta}^2}  - \frac{\delta \lambda^{Z\alpha} (1 + s_3) s_1}{2 \, \tilde{\lambda}} \right) y \, x^2 \right]$ \\
$\pm 0-+$ &
$ \frac{- (1\pm\cos\theta)\, \sqrt{\tilde{\lambda}}}{\sqrt{2s_3}} \left[\frac{x^2}{c_{\hat{\theta}}^2}
+\frac{y \, s_3  \,\delta \lambda^{Z\alpha}}{2}
+ \frac{y \, x^2}{2}
\left(\delta g_{1}^{Z\alpha} -2 \delta F_2^{Z\alpha}  + \delta \kappa^{Z\alpha} + \frac{s_3 \delta \lambda_Z}{c_{\hat\theta}^2}
- \frac{s_1 \, s_3 \, (1+ s_3)}{(1- s_3)^2}\,\delta \lambda^{Z\alpha}  \right) \right]
$\\
$0\pm-+$ &
$  \frac{(1\mp\cos\theta) \sqrt{\tilde{\lambda}} x}{\sqrt{2s_1}} \left[\frac{1}{c_{\hat{\theta}}^2}
  + \frac{y}{2}
 \left(\delta g_{1}^{Z\alpha} -2 \delta F_2^{Z\alpha} + \delta \kappa^{Z\alpha} + s_1 \,\delta \lambda^{Z\alpha} \right) \right]$
\\
$00+-$ & $ - \frac{\sin \theta}{\sqrt{s_1 \, s_3} \,  \sqrt{\tilde{\lambda}} \, s_{\hat\theta}^2} \, \frac{s_3 (s_3^2-1)}{4 \, x}$\\
$\pm\pm+-$ & $
\frac{\sin \theta \, \sqrt{\tilde{\lambda}}}{2 \, s_{\hat{\theta}}^2}
\left[1 - \frac{1}{\tilde{\lambda}}\left(1   - \frac{s_3 (1 - s_3  \pm s_3
\sqrt{\tilde{\lambda}})}{1-s_3 + \sqrt{\tilde{\lambda}} \cos \theta}\right)
- s_{\hat{\theta}}^2  F_3(\lambda_{\alpha},\lambda_{Z}) y\right]$ \\
$\pm 0+-$ & $
			\frac{-(1\mp\cos\theta) \sqrt{\tilde{\lambda}}}{2 \sqrt{2s_3} \, s_{\hat{\theta}}^2}
      \left[1 - \frac{1}{\tilde{\lambda}}\left(1  + s_3  - \frac{2 \, s_3 (1 - s_3  \pm
      \sqrt{\tilde{\lambda}})}{1-s_3 + \sqrt{\tilde{\lambda}} \cos \theta}\right)
      - s_{\hat{\theta}}^2 s_3  F_3(\lambda_{\alpha},\lambda_{Z}) y \right]$
\\
$0\pm +-$ &
$\frac{s_3 (s_3 - 1)(1 \pm \cos \theta)}{2 \, \sqrt{2 \, s_1} \, s_{\hat{\theta}}^2 \,  \sqrt{\tilde{\lambda}} \, x} \, \frac{1}{x}$

\\
$\pm\mp+-$ & $
		\frac{(\mp1+ \cos\theta)\sin\theta}
		{2 \, s_{\hat\theta}^2\left(1 - s_3 +\sqrt{\tilde{\lambda}} \, \cos\theta\right)}
		$ \\

\hline\end{tabular}
\caption{Expansion in $x,y < 1$ for the off-shell region of phase space of the CC03 diagrams and
$\tilde{\lambda} = \tilde{\lambda}(0,s_3)$. \label{Table:partialonoffshellb}}
\end{table}

These results make clear that non-interference arguments based on on-shell simplifications of the kinematics of decaying $W^\pm$ bosons
get off-shell corrections
for an LHC observable that includes
off-shell intermediate  $W^\pm$ kinematics. (Admittedly a somewhat obvious result.)
Such kinematics are parametrically suppressed by the small
width of the unstable gauge boson, but are generically included in LHC observables due to realistic experimental cuts.\footnote{In some cases, off-shell effects are not relevant for physical conclusions.
For example, Ref.~\cite{Cheung:2015aba} used helicity arguments similar to those employed here to
study the approximate holomorphy of the anomalous dimension matrix
of the SMEFT \cite{Alonso:2014rga}. Ref.~\cite{Cheung:2015aba} was focused on the cut-constructable part
of the amplitude related to logarithmic terms and the corresponding
divergences. As noted in Ref.~\cite{Cheung:2015aba} such reasoning does not not apply to finite contributions, which can
come about due to off-shell effects.}

\section{Mapping to past results} \label{sec:mapping}
The results in Table \ref{Table:on-shell},\ref{Table:bothoffshell},\ref{Table:partialonoffshella},\ref{Table:partialonoffshellb} are input parameter scheme independent,
and can be applied to more than one basis for $\mathcal{L}^{(6)}$.
Specializing to the Warsaw basis of operators, and the electroweak input
parameter scheme $\{\hat{\alpha}_{ew}, \hat{m}_Z, \hat{G}_F \}$ the (re-scaled) $x^2 y \, \delta X$ parameters are given by
\begin{subequations}
\begin{align*}
\frac{\hat{m}_W^2}{\Lambda^2} \, \delta g_1^{\alpha}  &= 0,  &
\frac{\hat{m}_W^2}{\Lambda^2} \, \delta \kappa_\alpha &=   \frac{1}{ \sqrt{2}\hat{G}_F}\frac{c_{\that}}{s_{\that}} C_{HWB},\\
\frac{\hat{m}_W^2}{\Lambda^2} \, \delta \lambda_{\alpha} &=  6 s_{\that}  \frac{\hat{m}^2_W}{\sqrt{4 \, \pi \, \hat{\alpha} }}C_W,  &
\frac{\hat{m}_W^2}{\Lambda^2} \, \delta \lambda_{Z} &=  6 s_{\that}  \frac{\hat{m}^2_W}{\sqrt{4 \, \pi \, \hat{\alpha} }}C_W,\\
\frac{\hat{m}_W^2}{\Lambda^2} \, \delta F_{1,2}^{\alpha} &= 0, &
\end{align*}
\end{subequations}
and
\begin{subequations}
\begin{align*}
-\frac{\hat{m}_W^2}{\Lambda^2} \, \frac{\delta F_1^{Z}}{4 \pi \hat{\alpha}} &= \delta \bar{g}_Z \, (g^{\ell}_{L})^{SM}_{ss} - \frac{1}{2 \sqrt{2} \hat{G}_F} \left(C_{\substack{H \ell \\ ss}}^{(1)} + C_{\substack{H \ell \\ ss}}^{(3)} \right) - \delta s_\theta^2, \\
-\frac{\hat{m}_W^2}{\Lambda^2} \, \frac{\delta F_2^{Z}}{4 \pi \hat{\alpha}} &= \delta \bar{g}_Z \, (g^{\ell}_{R})^{SM}_{ss}  - \frac{1}{2 \, \sqrt{2} \, \hat{G}_F}
C_{\substack{H e \\ss}} - \delta s_\theta^2,\\
\frac{\hat{m}_W^2}{\Lambda^2} \, \delta g_1^Z &=  \frac{1}{2 \sqrt{2}\hat{G}_F}\left(\frac{s_\that}{c_{\that}}+\frac{c_{\that}}{s_{\that}}  \right) C_{HWB} +
\frac{1}{2}\delta s_{\theta}^2\left(\frac{1}{s_{\that}^2}+\frac{1}{c_{\that}^2}\right),  \\
\frac{\hat{m}_W^2}{\Lambda^2} \, \delta \kappa_Z \, &=\frac{1}{ 2 \sqrt{2}\hat{G}_F}\left(- \frac{s_\that}{c_{\that}}+\frac{c_{\that}}{s_{\that}}  \right)C_{HWB} +
 \frac{1}{2}\delta s_{\theta}^2\left(\frac{1}{s_{\that}^2}+\frac{1}{c_{\that}^2}\right),
\end{align*}
\end{subequations}
with $\delta \bar{g}_Z,\delta s_\theta^2$ defined in the Appendix.
The left and right handed couplings are $(g^{\ell}_{L})^{SM}_{ss} = -1/2 + s_{\that}^2$, and $(g^{\ell}_{R})^{SM}_{ss} = s_{\that}^2$.
Here $s= \{1,2,3\}$ is a flavour index labeling the initial state leptons.
The results in Table \ref{Table:on-shell} can be more directly compared to Refs.~\cite{Azatov:2016sqh,Falkowski:2016cxu,Azatov:2017kzw,Panico:2017frx,Baglio:2017bfe}
using this procedure, finding agreement in the subset of terms
that were reported in these works. This comparison also utilizes the naive narrow width limit to simplify the amplitudes
as follows. In the sense of a distribution over phase space, the following replacement is made
\bea\label{narrowwidth}
|\bar D_W(s_{12})\bar D_W(s_{34})|^2 \, d s_{12} \, d s_{34} \rightarrow \frac{\pi^2}{\bar{m}_W^2 \, \bar{\Gamma}_W^2} \, \delta(s_{12} - \bar{m}_W^2)
\, \delta(s_{34} - \bar{m}_W^2) \, d s_{12} \, d s_{34}.
\eea
The result of this replacement is a factorizing of the diboson production mechanism $d \sigma(\bar{\psi} \psi \rightarrow W^+ \, W^-)/d\Omega$
and the branching ratios of the $W^\pm$ decays into specified final states as $s_1 = s_3 = 1$ is fixed in Table \ref{Table:on-shell}.
This approximation holds up to $\mathcal{O}(\Gamma_W/M_W)$ corrections to Eqn.~\ref{narrowwidth}.
The corrections in Tables
\ref{Table:bothoffshell},\ref{Table:partialonoffshella},\ref{Table:partialonoffshellb} are present and should not be overlooked by the construction
of a simplified high energy expansion, that is formally unphysical. It is not advisable to extrapolate the limited phase space results of Table \ref{Table:on-shell}
to the full phase space.

Another key difference between more recent studies of interference in the SMEFT in the high
energy limit, compared to the past studies of
interference of higher dimensional operators in the high energy limit for gluonic operators
\cite{Simmons:1989zs,Dixon:1993xd}, is the presence of an unstable massive gauge boson.
Such massive gauge bosons have been studied using the narrow width approximation.
However, a too naive version of the narrow width approximation does not commute with the SMEFT expansion.

This non-commutation can be seen as follows.
Expanding the propagator of the intermediate $W$ boson in the SMEFT
\begin{eqnarray}
  \frac{1}{(p^2-\bar m_W^2)^2+\bar\Gamma_W^2\bar m_W^2} =
  \frac{1}{(p^2-\hat m_W^2)^2+\hat \Gamma_W^2\hat m_W^2}
  (1 + \delta D_{W}(p^2) +\delta D_{W}(p^2)^*)
\end{eqnarray}
one has
\begin{eqnarray}
  \delta D_W(p^2)
  = \frac{1}{p^2- \hat m_W^2 + i \hat \Gamma_W \hat m_W} \times
  \left[\left(1-\frac{i\hat\Gamma_W}{2\hat m_W}\right) \delta m_W^2
  - i\hat m_W \delta \Gamma_W\right].
\end{eqnarray}
By first doing the narrow width approximation, and then doing the SMEFT expansion,
one obtains
\bea
\label{eq:nwaSMEFT}
  \frac{dp^2}{(p^2-\bar m_W^2)^2+\bar\Gamma_W^2\bar m_W^2}
  &\rightarrow& \frac{\pi dp^2}{\bar \Gamma_W \bar m_W}\delta(p^2-\bar m_W^2)
  \nonumber \\
  &=& \frac{\pi dp^2}{\hat\Gamma_W\hat m_W}\left(1 - \frac{\delta m_W^2}{2\hat m_W^2}
  - \frac{\delta\Gamma_W}{\hat\Gamma_W}\right)\delta(p^2 - \bar m_W^2).
\eea
Reversing the order of operations, we square the expanded propagators and then
do the narrow width approximation. For a general function $f(p^2)$, we find that after integrating
\bea
\label{eq:SMEFTnwa}
  \frac{f(p^2)dp^2}{(p^2-\hat m_W^2)^2 + \hat \Gamma_W^2\hat m_W^2} &\,& \hspace{-0.5cm}
  \left(1 + \delta D_W(p^2) + \delta D_W(p^2)^*\right) \nonumber \\
&\rightarrow& \frac{f(\hat m_W^2)\pi}{\hat\Gamma_W\hat m_W}\left(1-
  \frac{\delta m_W^2}{2\hat m_W^2}
  - \frac{\delta\Gamma_W}{\hat\Gamma_W}\right)+ \frac{f'(\hat m_W^2)\pi}{\hat\Gamma_W \hat m_W}\delta m_W^2.
\eea
In a naive version of the narrow width approximation, we simply replace $\bar m_W$ by
	$\hat m_W$ in Eqn.~\eqref{eq:nwaSMEFT}.
	The operations of expanding in the SMEFT and doing the naive narrow width
	approximation don't commute in general. The reason is that the naive narrow width approximation
assumes that the part of the integrand that is odd in its dependence on the
invariant mass cancels out in the near on-shell
region. With the SMEFT corrections, this is no longer the case, as the real part
of $\delta D_W$ gives a finite contribution to this part of the integrand.
This difference is proportional to the shift of the mass of the $W^\pm$ boson.
The correct way to implement the narrow width approximantion in the SMEFT is
	to use Eqn.~\eqref{eq:nwaSMEFT} and expand the general fuction $f(p^2)$ in
	the SMEFT expansion after integration. We then obtain Eqn.~\eqref{eq:SMEFTnwa}, and
see that the commutation property is restored.
Furthermore, we note that the $x$ expansion parameter itself can be chosen to be
$\hat{m}_W/\sqrt{s}$ or $\bar{m}_W/\sqrt{s}$ when studying the high energy limit (we choose the former expansion parameter).
This is another ambiguity that can be introduced into studies of this form, when using a
$\{\hat \alpha, \hat m_Z, \hat G_F\}$ scheme.

\section{Single charge current resonant contributions  (CC11)} \label{sec:singleresonant}
\begin{figure}[t]
	\centering
	\begin{subfigure}{0.20 \textwidth}
	\includegraphics[height=3.5cm,trim={ 0.5cm 0.0cm 0.0cm 0.0cm }, clip]{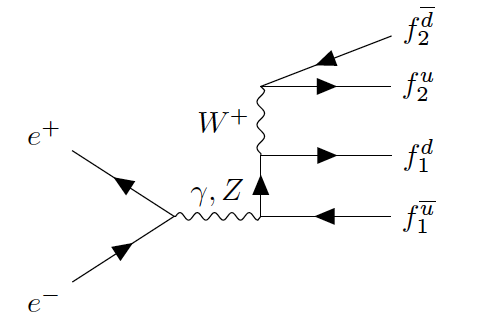}
	\label{fig:CC11achannel}
	\end{subfigure}
	\begin{subfigure}{0.40 \textwidth} \hspace{2cm}
	\includegraphics[height=3.5cm,trim={ 0.5cm 0cm 0cm 0cm }, clip]{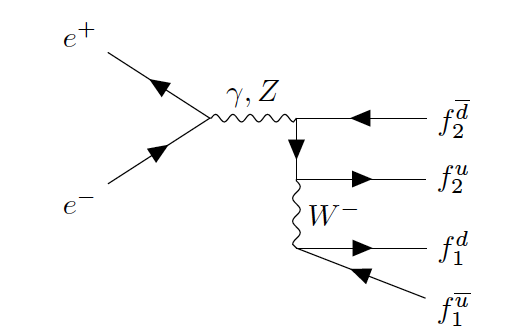}
	\label{fig:CC11bchannel}
	\end{subfigure}
	\caption{A subset of the CC11 Feynman diagrams contributing to $\bar{\psi} \psi \rightarrow \bar{\psi'}_1 \psi_2' \bar{\psi'}_3 \psi'_4$ with leptonic initial states.}
	\label{fig:CC11}
\end{figure}

It is well known in the SM literature, that the CC03 diagrams, with $W^\pm$ bosons fixed to be on-shell, are an insufficient approximation to a
$\bar{\psi} \psi \rightarrow \bar{\psi'}_1 \psi_2' \bar{\psi'}_3 \psi'_4$ cross section
  to describe the full phase space of scattering events \cite{Berends:1994pv,Denner:1999gp,Argyres:1995ym,Beenakker:1996kt,Beenakker:1996kn,Grunewald:2000ju,Grazzini:2016ctr}.
Such scattering events need not proceed through the CC03 set of diagrams, so limiting an analysis to this
set of diagrams is formally unphysical. This issue can be overcome using the standard techniques of expanding
around the poles of the process \cite{moellerpole,Stuart:1991xk,Veltman:1992tm}
and including more contributions to the physical scattering process due to single resonant or non-resonant diagrams.
Including the effect of single resonant diagrams allows one to develop gauge invariant
results for such scattering events \cite{Berends:1994pv,Denner:1999gp,Argyres:1995ym,Beenakker:1996kt,Beenakker:1996kn,Grunewald:2000ju}
when considering the full phase space (so long as the initial and final states are distinct).
Including the single resonant diagrams is frequently referred to as
calculating the set of CC11 diagrams in the literature.
Some of the additional diagrams required are
shown in Fig.~\ref{fig:CC11}.\footnote{Note that the CC03 diagrams are a (gauge dependent)
subset of the CC11 diagrams \cite{Beenakker:1994vn} which can be seen considering the differences found in CC03 results comparing axial
and $R_\xi$ gauges.}

Considering the results in the
previous sections, it is of interest to check if single resonant diagrams
contribute to the physical $\bar{\psi} \psi \rightarrow \bar{\psi'}_1 \psi_2' \bar{\psi'}_3 \psi'_4$ observables
in a manner that potentially cancels the contributions for the off-shell phase
space results in Tables \ref{Table:bothoffshell},\ref{Table:partialonoffshella},\ref{Table:partialonoffshellb}.
We find this is not the case, as can be argued on general grounds, and demonstrated in explicit calculations
which we report below.

In general, an expansion of a SM Lagrangian parameter with a SMEFT correction
is generically considered to be a correction of the form
\bea\label{SMshift}
\bar{X} = \hat{X} + x^2 \, y \, \delta X
\eea
in the high energy limit considered, and one expects the SMEFT shifts to enter at two higher orders in the $x$ expansion
compared to a SM result.
In addition the SMEFT can introduce new operator forms that directly lead to high energy growth and scale
as a $y$ correction to the amplitude, such as the effect of the operator $\mathcal{Q}_W$ in
$\bar{\psi} \psi \rightarrow \bar{\psi'}_1 \psi_2' \bar{\psi'}_3 \psi'_4$ scattering.

The CC03 diagram results are quite unusual due to the intricate cancellation present
between the leading terms in the $x$ expansion in the SM, at least in some regions of phase space. This leads to
the SM and SMEFT terms occurring in some cases at the same order in $x$,
contrary to the expectation formed by Eqn.~\ref{SMshift}. Conversely, the CC11 diagram contributions\footnote{
Modulo the CC03 diagrams which we indicate
with $\text{CC}11/\text{CC}03$.} follow the expectation in Eqn.~\ref{SMshift}.

\subsection{Single charge current resonant contributions - the SM}
We use the results of Refs.~\cite{Berends:1994pv,Denner:1999gp,Argyres:1995ym,Beenakker:1996kt,Beenakker:1996kn,Grunewald:2000ju},
in particular Ref.~\cite{Denner:1999gp}, for the SM results of the $\text{CC}11/\text{CC}03$ diagrams.
We neglect contributions suppressed by light fermion masses.
The generic SM amplitude is defined to have the form
\bea
i \, \mathcal{M}_{V_1 \, V_2}^{\sigma_a \, \sigma_b \, \sigma_c \, \sigma_d \, \sigma_e \, \sigma_f}(p_a,p_b,p_c,p_d,p_e,p_f)
&=& - 4 i  \, \bar{e}^4 \, \delta_{\sigma_a,-\sigma_b}\delta_{\sigma_c,-\sigma_d}
\delta_{\sigma_e,-\sigma_f} \bar{g}_{V_1\bar f_a f_g}^{\sigma_b}
\bar{g}_{V_2\bar f_g f_b}^{\sigma_b} \bar{g}_{V_1\bar f_c f_d}^{\sigma_d} \bar{g}_{V_2\bar f_e f_f}^{\sigma_f}
\nonumber\\
&\,&  \hspace{-1cm} \times \frac{\bar{D}_{V_1}(p_c+p_d)\bar{D}_{V_2}(p_e+p_f)}{(p_b+p_e+p_f)^2}A_2^{\sigma_a,\sigma_c,\sigma_e}
(p_a,p_b,p_c,p_d,p_e,p_f). \nonumber\\
\eea
We have adopted the conventions of Ref.~\cite{Denner:1999gp}, and the initial and final states are
labelled as $a \, b \rightarrow c \, d \, e\, f$. See the Appendix for more notational
details.
The functions $A_2^{\sigma_a,\sigma_c,\sigma_e}$ are given in terms of
spinor products as \cite{Denner:1999gp,Dittmaier:1998nn},
\bea
A_2^{+++}(p_a,p_b,p_c,p_d,p_e,p_f) &=&\langle p_ap_c\rangle
   \langle p_bp_f\rangle^* \left(\langle p_bp_d\rangle^* \langle p_bp_e\rangle
  + \langle p_dp_f\rangle^* \langle p_ep_f\rangle\right), \nonumber \\
\eea
and satisfy \cite{Denner:1999gp,Dittmaier:1998nn}
\bea
A_2^{++-}(p_a,p_b,p_c,p_d,p_e,p_f) &=& A_2^{+++}(p_a,p_b,p_c,p_d,p_f,p_e), \\
A_2^{+-+}(p_a,p_b,p_c,p_d,p_e,p_f) &=& A_2^{+++}(p_a,p_b,p_d,p_c,p_e,p_f), \\
A_2^{+--}(p_a,p_b,p_c,p_d,p_e,p_f) &=& A_2^{+++}(p_a,p_b,p_d,p_c,p_f,p_e), \\
A_2^{-,\sigma_c,\sigma_d}(p_a,p_b,p_c,p_d,p_e,p_f)
&=& \left(A_2^{+,-\sigma_c,-\sigma_d}(p_a,p_b,p_c,p_d,p_e,p_f)\right)^*.
\eea
The $\text{CC}11/\text{CC}03$ results are
		\begin{align}
			\mathcal{M}^{\sigma_+,\sigma_-,\sigma_1,\sigma_2,\sigma_3,\sigma_4}
			= \sum_{V=A,Z}\left[ \right. &\mathcal{M}_{VW}^{-\sigma_1,-\sigma_2,\sigma_+,\sigma_-,-\sigma_3,-\sigma_4}
			(-k_1,-k_2,p_+,p_-,-k_3,-k_4), \nonumber \\ +
			&\mathcal{M}_{VW}^{-\sigma_3,-\sigma_4,\sigma_+,\sigma_-,-\sigma_1,-\sigma_2}
			(-k_3,-k_4,p_+,p_-,-k_1,-k_2), \nonumber \\+
			&\mathcal{M}_{WV}^{-\sigma_1,-\sigma_2,-\sigma_3,-\sigma_4,\sigma_+,\sigma_-}
			(-k_1,-k_2,-k_3,-k_4,p_+,p_-), \nonumber \\+
			&\mathcal{M}_{WV}^{-\sigma_3,-\sigma_4,-\sigma_1,-\sigma_2,\sigma_+,\sigma_-}
			(-k_3,-k_4,-k_1,-k_2,p_+,p_-) \left.\right].
		\end{align}
As the final state fermions couple to one $W^\pm$ boson, and fermion masses are neglected,
$\{\sigma_1,\sigma_2,\sigma_3,\sigma_4\}=\{-+-+\}$. We denote the amplitude
by the helicities of the incoming fermions,
$\mathcal{M}^{\sigma_+,\sigma_-,\sigma_1,\sigma_2,\sigma_3,\sigma_4}
=\mathcal{M}^{\sigma_+,\sigma_-}$ and find using \cite{Denner:1999gp} in the $x<1$ limit
for Case 1 and right handed electrons
\begin{align}
  \mathcal{M}^{-+}= \frac{\hat{e}^4Q_l\sin\theta\sin\tilde\theta_{12}
  \sin\tilde\theta_{34}}{4s_{\hat{\theta}}^2 \, c_{\hat{\theta}}^2 \, s \, x^2}
    \left[\frac{Q_{f_1}-I^3_{f_1}-Q_{f_2}+I^3_{f_2}}
    {s_3-1+i \hat{\gamma}_W}
    +\frac{Q_{f_4}-I^3_{f_4}
    -Q_{f_3}+I^3_{f_3}}
  {s_1-1+i \hat{\gamma}_W}\right],
\end{align}
and for left handed electrons
\begin{align}
  \mathcal{M}^{+-}=& \frac{\hat{e}^4\sin\theta\sin\tilde\theta_{12}
  \sin\tilde\theta_{34}}{4s_{\hat{\theta}}^4 \, c_{\hat{\theta}}^2 \, s \, x^2}
  \nonumber \\
  &\left[\frac{[Q_{f_1}s_{\hat{\theta}}^2(Q_l - I^3_l)+I_{f_1}^3(I^3_l-Q_l \, s_{\hat{\theta}}^2)]
    -[Q_{f_2}\, s_{\hat{\theta}}^2(Q_l-I^3_l)+I_{f_2}^3(I^3_l-Q_l\, s_{\hat{\theta}}^2)] }{s_3-1+i\hat{\gamma}_W}\right.
    \nonumber \\
    &+\left.\frac{Q_{f_4}s_{\hat{\theta}}^2(Q_l - I^3_l)+I_{f_4}^3(I^3_l-Q_l \, s_{\hat{\theta}}^2)]
    -[Q_{f_3}\, s_{\hat{\theta}}^2(Q_l-I^3_l)+I_{f_3}^3(I^3_l-Q_l\, s_{\hat{\theta}}^2)] }{s_1-1+i\hat{\gamma}_W}\right].
\end{align}
Here $\hat{\gamma}_W = \hat{\Gamma}_W/\hat{m}_W$, $Q_{f_i}$ is the electric charge and $I_{f_i}^3 = \pm 1/2$
is the isospin of the fermion $f_i$.
Similarly for Case 2 we find using \cite{Denner:1999gp} the results for right handed electrons
\begin{align}\label{case2rights}
  \mathcal{M}^{-+}=\frac{4\hat{e}^4 Q_l}{s_{\hat{\theta}}^2c_{\hat{\theta}}^2 s}
  &\left[\frac{I^3_{f_1}-Q_{f_1}}{s_3(1-s_1+s_3-\tilde\lambda\cos\tilde\theta_{12})}R_1 - \frac{I^3_{f_2}-Q_{f_2}}{s_3(1-s_1+s_3+\tilde\lambda\cos\tilde\theta_{12})}R_2\right.
  \nonumber \\&+
  \left.\frac{I^3_{f_3}-Q_{f_3}}{s_1(1+s_1-s_3-\tilde\lambda\cos\tilde\theta_{34})}R_3 - \frac{I^3_{f_4}-Q_{f_4}}{s_1(1+s_1-s_3+\tilde\lambda\cos\tilde\theta_{34})}R_4\right],
\end{align}
and for left handed electrons
\begin{align}
  \mathcal{M}^{+-}
=\frac{-4 \hat{e}^4}{s_{\hat{\theta}}^4 c_{\hat{\theta}}^2 s}
  &\left[\frac{Q_{f_1}s_{\hat{\theta}}^2(Q_l-I^3_l) +I^3_{f_1}(I^3_l-Q_ls_{\hat{\theta}}^2)}{s_3(1-s_1+s_3-\tilde\lambda\cos\tilde\theta_{12})}L_1 \right.
  - \frac{Q_{f_2}s_{\hat{\theta}}^2(Q_l-I^3_l) +I^3_{f_2}(I^3_l-Q_ls_{\hat{\theta}}^2)}{s_3(1-s_1+s_3+\tilde\lambda\cos\tilde\theta_{12})}L_2 \nonumber  \\
  &+ \left.
\frac{Q_{f_3}s_{\hat{\theta}}^2(Q_l-I^3_l) +I^3_{f_3}(I^3_l-Q_ls_{\hat{\theta}}^2)}{s_1(1+s_1-s_3-\tilde\lambda\cos\tilde\theta_{34})}L_3
  -  \frac{Q_{f_4}s_{\hat{\theta}}^2(Q_l-I^3_l) +I^3_{f_4}(I^3_l-Q_ls_{\hat{\theta}}^2)}{s_1(1+s_1-s_3+\tilde\lambda\cos\tilde\theta_{34})}L_4\right]. \label{case2lefts}
\end{align}
The functions $R_i$,$L_i$, $i=1,..,4$ are given in the Appendix, along with additional
definitions. For Case 3a one finds for right handed electrons
\bea\label{case3arights}
    \mathcal{M}^{-+}&=&\frac{\hat{e}^4Q_l\sin\tilde\theta_{34}}
  {4s_{\hat{\theta}}^2c_{\hat{\theta}}^2 sx^2(s_3-1+i\gamma_W)} \left[ (Q_{f_1}-I^3_{f_1}) -(Q_{f_2}-I^3_{f_2}) \right] \times
  \\ &\,& \hspace{-1cm}
  \left(\sin\theta\sin\tilde\theta_{12}(1+s_1)
+\sqrt{s_1}e^{-i\tilde\phi_{12}}(1-\cos\theta)(1+\cos\tilde\theta_{12})
+\sqrt{s_1}e^{i\tilde\phi_{12}}(1+\cos\theta)(1-\cos\tilde\theta_{12})\right), \nonumber
\eea
and for left-handed electrons
\bea
\mathcal{M}^{+-}&=&
\frac{\hat{e}^4\sin\tilde\theta_{34}}
  {4s_{\hat{\theta}}^4c_{\hat{\theta}}^2 s x^2(s_3-1+i\gamma_W)}\times \\
  &\,&\left[(Q_{f_1}s_{\hat{\theta}}^2(Q_l-I^3_l)+
    I^3_{f_1}(I^3_l-Q_ls_{\hat{\theta}}^2))
    -(Q_{f_2}s_{\hat{\theta}}^2(Q_l-I^3_l)+
  I^3_{f_2}(I^3_l-Q_ls_{\hat{\theta}}^2))\right] \times \nonumber
  \\ &\,& \hspace{-1cm}\left[
  \sin\theta\sin\tilde\theta_{12}(1+s_1)
-\sqrt{s_1}e^{-i\tilde\phi_{12}}(1+\cos\theta)(1+\cos\tilde\theta_{12})
-\sqrt{s_1}e^{i\tilde\phi_{12}}(1-\cos\theta)(1-\cos\tilde\theta_{12})\right], \nonumber
\eea
and finally for Case 3b one finds for right-handed electrons
\bea
  \mathcal{M}^{-+}&=&\frac{\hat{e}^4Q_l\sin\tilde\theta_{12}}
  {4s_{\hat{\theta}}^2c_{\hat{\theta}}^2 sx^2(s_1-1+i\gamma_W)}\left[(Q_{f_4}-I^3_{f_4}) -(Q_{f_3}-I^3_{f_3})\right]\times \\
   &\,& \hspace{-1cm}
  \left(\sin\theta\sin\tilde\theta_{34}(1+s_3)
-\sqrt{s_3}e^{-i\tilde\phi_{34}}(1-\cos\theta)(1-\cos\tilde\theta_{34})-\sqrt{s_3}e^{i\tilde\phi_{34}}(1+\cos\theta)(1+\cos\tilde\theta_{34})\right), \nonumber
\eea
and for left-handed electrons
\bea
  \mathcal{M}^{+-}&=&\frac{\hat{e}^4\sin\tilde\theta_{12}}
  {4s_{\hat{\theta}}^4c_{\hat{\theta}}^2 sx^2(s_1-1+i\gamma_W)} \times \label{case3blefts} \\
  &\,&\left[(Q_{f_4}s_{\hat{\theta}}^2(I^3_l-Q_l s_{\hat{\theta}}^2)+
    I^3_{f_4}(I^3_l-Q_ls_{\hat{\theta}}^2))
    -(Q_{f_3}s_{\hat{\theta}}^2(I^3_l-Q_ls_{\hat{\theta}}^2)+
  I^3_{f_3}(I^3_l-Q_ls_{\hat{\theta}}^2))\right] \times  \nonumber
  \\ &\,& \hspace{-1cm}\left[
  \sin\theta\sin\tilde\theta_{34}(1+s_3)
+\sqrt{s_3}e^{-i\tilde\phi_{34}}(1+\cos\theta)(1-\cos\tilde\theta_{34})
+\sqrt{s_3}e^{i\tilde\phi_{34}}(1-\cos\theta)(1+\cos\tilde\theta_{34})\right]. \nonumber
\eea
\subsection{Single resonant contributions - the SMEFT}

The SMEFT corrections to the single resonant charged current contributions to  $\bar{\psi} \psi \rightarrow \bar{\psi'}_1 \psi_2' \bar{\psi'}_3 \psi'_4$,
follow directly from the results in the previous section. These corrections follow the scaling in
$x$ expectation formed by Eqn.~\ref{SMshift}, and the spinor products are unaffected by these shifts.
As the charges of the initial and final states through neutral currents are fairly explicit in the previous section,
it is easy to determine the coupling shifts and the SMEFT corrections to the
propagators ($\delta D_{W,Z}$) by direct substitution.

We find that the single resonant contributions are distinct in their kinematic dependence
compared to the novel interference
results we have reported in Section \ref{sec:on-shell}.
The direct comparison of the results is somewhat challenged by the lack
of a meaningful decomposition of the single resonant diagrams into helicity eigenstates
of two intermediate charged currents, when only one charged current is present.
Furthermore, we also note that the angular dependence shown in the single resonant
results in Eqns.~\ref{case2rights}-\ref{case3blefts} reflects the fact that
the center of mass frame relation to the final state phase space
in the case of the CC03 diagrams is distinct from the CC11/CC03 results. This is the
case despite both contributions being required
for gauge independence in general \cite{Beenakker:1994vn}.

\begin{figure}[t]
	\centering
	\begin{subfigure}{0.30 \textwidth}
	\includegraphics[height=3.5cm,trim={ 0.5cm 0.0cm 0.0cm 0.0cm }, clip]{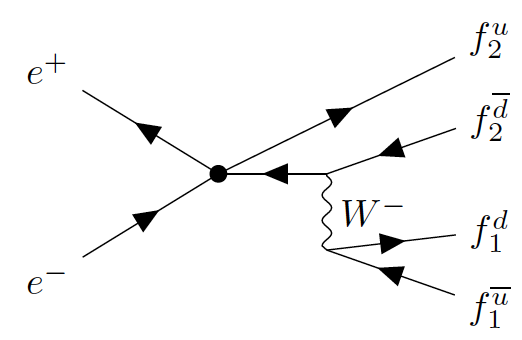}
	\label{fig:FFcase1}
	\end{subfigure}
	\begin{subfigure}{0.40 \textwidth} \hspace{1cm}
	\includegraphics[height=3.5cm,trim={ 0.5cm 0cm 0cm 0cm }, clip]{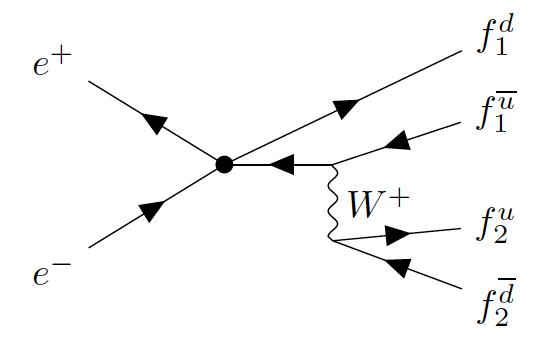}
	\label{fig:FFcase2}
	\end{subfigure}
	\caption{A subset of $\psi^4$ operator insertions contributing to
  $\bar{\psi} \psi \rightarrow \bar{\psi'}_1 \psi_2' \bar{\psi'}_3 \psi'_4$ scattering.}
	\label{fig:ff}
\end{figure}

To develop a complete SMEFT result including single resonant contributions,
it is also required to supplement the results in the previous section with four fermion diagrams
where a near on-shell charged current is present. For diagrams of this form see Fig.~\ref{fig:ff}.
These contributions introduce dependence on $\mathcal{L}^{(6)}$ operators that are not present in the CC03 diagrams,
and once again the angular dependence in the phase space is distinct from the
CC03 results.

\section{Conclusions}
In this paper, we have shown that off-shell effects in CC03 diagrams contributing to
$\bar{\psi} \psi \rightarrow \bar{\psi'}_1 \psi_2' \bar{\psi'}_3 \psi'_4$ observables lead to interference
between the SM and $\mathcal{L}^{(6)}$ operators in the high energy limit. These effects can be
overlooked when studying a simplified limit of these scattering events, as defined by
the CC03 diagrams and the narrow width approximation. We have determined the results of the CC03 diagrams in several novel
regions of phase space, compared to recent SMEFT literature, and have shown that single resonant diagrams do not change these conclusions
when included into the results.
We have also illustrated how to make the narrow width approximation consistent with
the SMEFT expansion.

The off-shell phase space of the CC03 diagrams considered, and the phase space of the single resonant
diagrams, is parametrically suppressed in an inclusive $\bar{\psi} \psi \rightarrow \bar{\psi'}_1 \psi_2' \bar{\psi'}_3 \psi'_4$  observable.
The full phase space is dominated by the near on-shell contributions of the CC03 diagrams which can be parametrically
larger by $\sim(\hat{\Gamma}_W \, \hat{m}_W/\bar{v}_T^2)^{-1}$
or $\sim (\hat{\Gamma}_W \, \hat{m}_W/p_i^2)^{-1}$ where $p_i^2$ is a Lorentz invariant of mass dimension two.
The exact degree of suppression that the off-shell region of phase space
experiences strongly depends on the experimental cuts defining the inclusive
$\bar{\psi} \psi \rightarrow \bar{\psi'}_1 \psi_2' \bar{\psi'}_3 \psi'_4$  observables, which should be studied
in a gauge independent manner including all diagrams that contribute to the experimental observable, i.e.
including all CC11 diagrams.

In some sense, our results coincide with the overall thrust of the discussion of Ref.~\cite{Azatov:2016sqh},
which emphasizes that searching for the effects of $\mathcal{L}^{(6)}$ operators
interfering with the SM in tails of distributions (i.e. in the
$\hat{m}_{W}^2/s \rightarrow 0$ limit) can be challenged in some helicity configurations,
by the smallness of such interference effects. Arguably, this encourages prioritizing
 SMEFT studies on ``pole observables'' and makes such LHC studies a higher priority
compared to pursuing such suppressed ``tail observables''.\footnote{For a recent discussion on a
 systematic SMEFT pole program see Ref.~\cite{Brivio:2017btx}. One of the comparative strengths of the pole
 observable program is that observables can be optimized so that interference suppression
 effects {\it enhance} theoretical control of a process for SMEFT studies.}
At the same time, we stress that the results of this work indicate that the strong statements
on non-interference of the SM and $\mathcal{L}^{(6)}$ operators, in subsets of phase space, and for some helicity configurations,
are tempered by finite width effects, in addition to perturbative corrections \cite{Dixon:1993xd,Azatov:2016sqh} and finite mass
suppressions \cite{Azatov:2016sqh}.
Finally, our results also demonstrate
that a careful examination of historical and rigorous SM results, in the well developed SM literature, are an essential
foundation to precise and accurate SMEFT studies.
\\
\\
\\
{\bf Acknowledgements}
We acknowledge generous support from the Villum Fonden and partial support by the Danish National Research Foundation (DNRF91).
We thank Yun Jiang, Giampiero Passarino, and Massimiliano Grazzini for helpful comments.

\appendix
\section{Conventions and notation}
We use the generic notation $\delta X = \bar{X} - \hat{X}$ for the
differences for a Lagrangian parameter $X$ \cite{Brivio:2017vri,Brivio:2017btx} due to $\mathcal{L}^{(6)}$
corrections in the SMEFT and define
\begin{align}
 \d G_F &= \frac{1}{\sqrt{2} \,  \hat{G}_F} \left(\sqrt{2} \, C^{(3)}_{\substack{Hl}} - \frac{C_{\substack{ll}}}{\sqrt{2}}\right), \\
 \d m_Z^2 &= \frac{1}{2 \, \sqrt{2}} \, \frac{\hat{m}_Z^2}{\hat{G}_F} C_{HD} + \frac{2^{1/4} \sqrt{\pi \hat{\alpha}} \, \hat{m}_Z}{\hat{G}_F^{3/2}} C_{HWB},\\
 \delta \bar{g}_Z &=- \frac{\delta G_F}{\sqrt{2}} - \frac{\delta m_Z^2}{2\hat{m}_Z^2} + \frac{s_{\hat{\theta}} \, c_{\hat{\theta}}}{\sqrt{2} \hat{G}_F} \, C_{HWB}, \\
\d g_1 &=\frac{\hat g_1}{2\hcdt}\left[\hst^2\left(\sqrt2 \d G_f+\frac{\d m_Z^2}{\hat m_Z^2}\right)+\hct^2\,\hsdt\bar v_T^2 C_{HWB}\right],\nonumber\\
 \d g_2 &=-\frac{\hat g_2}{2\hcdt}\left[
  \hct^2\left(\sqrt2 \d G_f+\frac{\d m_Z^2}{\hat m_Z^2}\right)+\hst^2\,\hsdt\bar v_T^2 C_{HWB}\right],\\  %
  \d \st^2 &= 2\hct^2\hst^2\left(\frac{\d g_1}{\hat g_1}-\frac{\d g_2}{\hat g_2}\right) +\bar v_T^2  \frac{\hsdt\hcdt}{2} C_{HWB}, \\
  \frac{\hat m_W^2}{\Lambda^2}\d \bar{g}_W^{\ell}&=\frac{1}{2\sqrt{2}\hat G_F}\left(C_{H\ell}^{(3)} + \frac{1}{2}\frac{c_{\hat\theta}}{s_{\hat\theta}}C_{HWB}\right) -\frac{1}{4}\frac{\d s_{\theta}^2}{s_{\hat\theta}^2}.
\end{align}

\begin{align}
  R_1 = \left[-\gamma_{12}^-e^{i\tilde\phi_{12}}\cos\frac{\theta}{2}\sin
    \frac{\tilde\theta_{12}}{2}-\sin\frac{\theta}{2}\cos\frac{\tilde
    \theta_{12}}{2}\right]
    \left[\gamma_{12}^-e^{i\tilde\phi_{34}}
      \cos\frac{\tilde\theta_{12}}{2}\cos
      \frac{\tilde\theta_{34}}{2}+\gamma_{34}^+
      e^{i\tilde\phi_{12}}\sin\frac{\tilde\theta_{12}}{2}
      \sin\frac{\tilde\theta_{34}}{2}\right] \nonumber \\
    \left\{\sqrt{s_1}\left[e^{-i\tilde\phi_{12}}
      \sin\frac{\theta}{2}\cos
      \frac{\tilde\theta_{12}}{2}+\gamma_{12}^+
      \cos\frac{\theta}{2}
      \sin\frac{\tilde\theta_{12}}{2}\right]
      \left[-\gamma_{12}^-e^{i\tilde\phi_{12}}
      \cos\frac{\tilde\theta_{12}}{2}\sin
      \frac{\tilde\theta_{34}}{2}+\gamma_{34}^+
      e^{i\tilde\phi_{34}}\sin\frac{\tilde\theta_{12}}{2}
      \cos\frac{\tilde\theta_{34}}{2}\right] \right. \nonumber \\
      \left.-\sqrt{s_3}\left[-e^{i\tilde\phi_{34}}
      \cos\frac{\theta}{2}\cos
      \frac{\tilde\theta_{34}}{2}+\gamma_{34}^+
     \sin\frac{\theta}{2}
    \sin\frac{\tilde\theta_{34}}{2}\right]\right\}e^{-i(\tilde\phi_{12}+\tilde\phi_{34})}\sqrt{s_1s_3}\gamma_{12}^+\gamma_{34}^-
\end{align}

\begin{align}
  R_2 = \left[\gamma_{12}^-\sin\frac{\theta}{2}\cos
    \frac{\tilde\theta_{12}}{2}+e^{i\tilde\phi_{12}}
    \cos\frac{\theta}{2}\sin\frac{\tilde
    \theta_{12}}{2}\right]
    \left[-\gamma_{12}^-e^{i\tilde\phi_{12}}
      \sin\frac{\tilde\theta_{12}}{2}\sin
      \frac{\tilde\theta_{34}}{2}-\gamma_{34}^+
      e^{i\tilde\phi_{34}}\cos\frac{\tilde\theta_{12}}{2}
      \cos\frac{\tilde\theta_{34}}{2}\right] \nonumber \\
    \left\{\sqrt{s_1}\left[
      \cos\frac{\theta}{2}\cos
      \frac{\tilde\theta_{12}}{2}-\gamma_{12}^+
      e^{-i\tilde\phi_{12}}
      \sin\frac{\theta}{2}
      \sin\frac{\tilde\theta_{12}}{2}\right]
      \left[\gamma_{12}^-e^{i\tilde\phi_{34}}
      \cos\frac{\tilde\theta_{12}}{2}\cos
      \frac{\tilde\theta_{34}}{2}+\gamma_{34}^+
      e^{i\tilde\phi_{12}}\sin\frac{\tilde\theta_{12}}{2}
      \sin\frac{\tilde\theta_{34}}{2}\right] \right. \nonumber \\
      \left.+\left[\gamma_{34}^+
      \sin\frac{\theta}{2}\sin
      \frac{\tilde\theta_{34}}{2}-e^{i\tilde\phi_{34}}
      \cos\frac{\theta}{2}
    \cos\frac{\tilde\theta_{34}}{2}\right]\right\}e^{-i(\tilde\phi_{12}+\tilde\phi_{34})}\sqrt{s_1s_3}\gamma_{12}^+\gamma_{34}^-
\end{align}
\begin{align}
  R_3 = \left[\gamma_{34}^+e^{i\tilde\phi_{34}}\cos\frac{\theta}{2}\cos
    \frac{\tilde\theta_{34}}{2}-\sin\frac{\theta}{2}\sin\frac{\tilde
    \theta_{34}}{2}\right]
    \left[\gamma_{12}^-e^{i\tilde\phi_{34}}
      \cos\frac{\tilde\theta_{12}}{2}\cos
      \frac{\tilde\theta_{34}}{2}+\gamma_{34}^+
      e^{i\tilde\phi_{12}}\sin\frac{\tilde\theta_{12}}{2}
      \sin\frac{\tilde\theta_{34}}{2}\right] \nonumber \\
    \left\{\sqrt{s_3}\left[e^{-i\tilde\phi_{34}}
      \sin\frac{\theta}{2}\sin
      \frac{\tilde\theta_{34}}{2}-\gamma_{34}^-
      \cos\frac{\theta}{2}
      \cos\frac{\tilde\theta_{34}}{2}\right]
      \left[-\gamma_{12}^-e^{i\tilde\phi_{12}}
      \sin\frac{\tilde\theta_{12}}{2}\cos
      \frac{\tilde\theta_{34}}{2}+\gamma_{34}^+
      e^{i\tilde\phi_{34}}\cos\frac{\tilde\theta_{12}}{2}
      \sin\frac{\tilde\theta_{34}}{2}\right] \right. \nonumber \\
      \left.+\sqrt{s_1}\left[\gamma_{12}^-
      \sin\frac{\theta}{2}\cos
      \frac{\tilde\theta_{12}}{2}+e^{i\tilde\phi_{12}}
      \cos\frac{\theta}{2}
    \sin\frac{\tilde\theta_{12}}{2}\right]\right\}e^{-i(\tilde\phi_{12}+\tilde\phi_{34})}\sqrt{s_1s_3}\gamma_{12}^+\gamma_{34}^-
\end{align}
\begin{align}
  R_4 = \left[\gamma_{34}^+\sin\frac{\theta}{2}\sin
    \frac{\tilde\theta_{34}}{2}
    -e^{i\tilde\phi_{34}}\cos\frac{\theta}{2}\cos\frac{\tilde
    \theta_{34}}{2}\right]
    \left[\gamma_{12}^-e^{i\tilde\phi_{12}}
      \sin\frac{\tilde\theta_{12}}{2}\sin
      \frac{\tilde\theta_{34}}{2}+\gamma_{34}^+
      e^{i\tilde\phi_{34}}\cos\frac{\tilde\theta_{12}}{2}
      \cos\frac{\tilde\theta_{34}}{2}\right] \nonumber \\
    \left\{\sqrt{s_3}\left[
      \cos\frac{\theta}{2}\sin
      \frac{\tilde\theta_{34}}{2}+\gamma_{34}^-
      e^{-i\tilde\phi_{34}}\sin\frac{\theta}{2}
      \cos\frac{\tilde\theta_{34}}{2}\right]
      \left[-\gamma_{12}^-e^{i\tilde\phi_{34}}
      \cos\frac{\tilde\theta_{12}}{2}\cos
      \frac{\tilde\theta_{34}}{2}-\gamma_{34}^+
      e^{i\tilde\phi_{12}}\sin\frac{\tilde\theta_{12}}{2}
      \sin\frac{\tilde\theta_{34}}{2}\right] \right. \nonumber \\
      \left.+\left[\gamma_{12}^-
      \sin\frac{\theta}{2}\cos
      \frac{\tilde\theta_{12}}{2}+e^{i\tilde\phi_{12}}
      \cos\frac{\theta}{2}
    \sin\frac{\tilde\theta_{12}}{2}\right]\right\}e^{-i(\tilde\phi_{12}+\tilde\phi_{34})}\sqrt{s_1s_3}\gamma_{12}^+\gamma_{34}^-
\end{align}
\begin{align}
  L_1 = \left[-\gamma_{12}^-e^{i\tilde\phi_{12}}\sin\frac{\theta}{2}\sin
    \frac{\tilde\theta_{12}}{2}+\cos\frac{\theta}{2}\cos\frac{\tilde
    \theta_{12}}{2}\right]
    \left[\gamma_{12}^-e^{i\tilde\phi_{34}}
      \cos\frac{\tilde\theta_{12}}{2}\cos
      \frac{\tilde\theta_{34}}{2}+\gamma_{34}^+
      e^{i\tilde\phi_{12}}\sin\frac{\tilde\theta_{12}}{2}
      \sin\frac{\tilde\theta_{34}}{2}\right] \nonumber \\
   \left\{ \sqrt{s_1}\left[e^{-i\tilde\phi_{12}}
      \cos\frac{\theta}{2}\cos
      \frac{\tilde\theta_{12}}{2}-\gamma_{12}^+
      \sin\frac{\theta}{2}\sin\frac{\tilde\theta_{12}}{2}\right]\right.
      \left[-\gamma_{12}^-e^{i\tilde\phi_{12}}
      \cos\frac{\tilde\theta_{12}}{2}\sin
      \frac{\tilde\theta_{34}}{2}+\gamma_{34}^+
      e^{i\tilde\phi_{34}}\sin\frac{\tilde\theta_{12}}{2}
     \cos\frac{\tilde\theta_{34}}{2}\right]
\nonumber \\
      \left.-\sqrt{s_3}\left[e^{i\tilde\phi_{34}}
      \sin\frac{\theta}{2}\cos
      \frac{\tilde\theta_{34}}{2}+\gamma_{34}^+
       \cos\frac{\theta}{2}
    \sin\frac{\tilde\theta_{34}}{2}\right]\right\}e^{-i(\tilde\phi_{12}+\tilde\phi_{34})}\sqrt{s_1s_3}\gamma_{12}^+\gamma_{34}^-
\end{align}

\begin{align}
  L_2 = \left[-\gamma_{12}^-\cos\frac{\theta}{2}\cos
    \frac{\tilde\theta_{12}}{2}+e^{i\tilde\phi_{12}}
    \sin\frac{\theta}{2}\sin\frac{\tilde
    \theta_{12}}{2}\right]
    \left[\gamma_{12}^-e^{i\tilde\phi_{12}}
      \sin\frac{\tilde\theta_{12}}{2}\sin
      \frac{\tilde\theta_{34}}{2}+\gamma_{34}^+
      e^{i\tilde\phi_{34}}\cos\frac{\tilde\theta_{12}}{2}
      \cos\frac{\tilde\theta_{34}}{2}\right] \nonumber \\
    \left\{\sqrt{s_1}\left[
      \sin\frac{\theta}{2}\cos
      \frac{\tilde\theta_{12}}{2}+\gamma_{12}^+
      e^{-i\tilde\phi_{12}}
      \cos\frac{\theta}{2}
      \sin\frac{\tilde\theta_{12}}{2}\right]
      \left[\gamma_{12}^-e^{i\tilde\phi_{34}}
      \cos\frac{\tilde\theta_{12}}{2}\cos
      \frac{\tilde\theta_{34}}{2}+\gamma_{34}^+
      e^{i\tilde\phi_{12}}\sin\frac{\tilde\theta_{12}}{2}
      \sin\frac{\tilde\theta_{34}}{2}\right] \right. \nonumber \\
      \left.-\left[\gamma_{34}^+
      \cos\frac{\theta}{2}\sin
      \frac{\tilde\theta_{34}}{2}+e^{i\tilde\phi_{34}}
      \sin\frac{\theta}{2}
    \cos\frac{\tilde\theta_{34}}{2}\right]\right\}e^{-i(\tilde\phi_{12}+\tilde\phi_{34})}\sqrt{s_1s_3}\gamma_{12}^+\gamma_{34}^-
\end{align}
\begin{align}
  L_3 = \left[\gamma_{34}^+e^{i\tilde\phi_{34}}\sin\frac{\theta}{2}\cos
    \frac{\tilde\theta_{34}}{2}+\cos\frac{\theta}{2}\sin\frac{\tilde
    \theta_{34}}{2}\right]
    \left[-\gamma_{12}^-e^{i\tilde\phi_{34}}
      \cos\frac{\tilde\theta_{12}}{2}\cos
      \frac{\tilde\theta_{34}}{2}-\gamma_{34}^+
      e^{i\tilde\phi_{12}}\sin\frac{\tilde\theta_{12}}{2}
      \sin\frac{\tilde\theta_{34}}{2}\right] \nonumber \\
    \left\{-\sqrt{s_3}\left[e^{-i\tilde\phi_{34}}
      \cos\frac{\theta}{2}\sin
      \frac{\tilde\theta_{34}}{2}+\gamma_{34}^-
      \sin\frac{\theta}{2}
      \cos\frac{\tilde\theta_{34}}{2}\right]
      \left[-\gamma_{12}^-e^{i\tilde\phi_{12}}
      \sin\frac{\tilde\theta_{12}}{2}\cos
      \frac{\tilde\theta_{34}}{2}+\gamma_{34}^+
      e^{i\tilde\phi_{34}}\cos\frac{\tilde\theta_{12}}{2}
      \sin\frac{\tilde\theta_{34}}{2}\right] \right. \nonumber \\
      \left.+\sqrt{s_1}\left[-\gamma_{12}^-
      \cos\frac{\theta}{2}\cos
      \frac{\tilde\theta_{12}}{2}+e^{i\tilde\phi_{12}}
      \sin\frac{\theta}{2}
    \sin\frac{\tilde\theta_{12}}{2}\right]\right\}e^{-i(\tilde\phi_{12}+\tilde\phi_{34})}\sqrt{s_1s_3}\gamma_{12}^+\gamma_{34}^-
    		\end{align}
\begin{align}
  L_4 = \left[\gamma_{34}^+\cos\frac{\theta}{2}\sin
    \frac{\tilde\theta_{34}}{2}
    +e^{i\tilde\phi_{34}}\sin\frac{\theta}{2}\cos\frac{\tilde
    \theta_{34}}{2}\right]
    \left[-\gamma_{12}^-e^{i\tilde\phi_{12}}
      \sin\frac{\tilde\theta_{12}}{2}\sin
      \frac{\tilde\theta_{34}}{2}-\gamma_{34}^+
      e^{i\tilde\phi_{34}}\cos\frac{\tilde\theta_{12}}{2}
      \cos\frac{\tilde\theta_{34}}{2}\right] \nonumber \\
    \left\{-\sqrt{s_3}\left[
      \sin\frac{\theta}{2}\sin
      \frac{\tilde\theta_{34}}{2}-\gamma_{34}^-
      e^{-i\tilde\phi_{34}}\cos\frac{\theta}{2}
      \cos\frac{\tilde\theta_{34}}{2}\right]
      \left[-\gamma_{12}^-e^{i\tilde\phi_{34}}
      \cos\frac{\tilde\theta_{12}}{2}\cos
      \frac{\tilde\theta_{34}}{2}-\gamma_{34}^+
      e^{i\tilde\phi_{12}}\sin\frac{\tilde\theta_{12}}{2}
      \sin\frac{\tilde\theta_{34}}{2}\right] \right. \nonumber \\
      \left.-\left[-\gamma_{12}^-
      \cos\frac{\theta}{2}\cos
      \frac{\tilde\theta_{12}}{2}+e^{i\tilde\phi_{12}}
      \sin\frac{\theta}{2}
    \sin\frac{\tilde\theta_{12}}{2}\right]\right\}e^{-i(\tilde\phi_{12}+\tilde\phi_{34})}\sqrt{s_1s_3}\gamma_{12}^+\gamma_{34}^-
\end{align}

\subsection{Phase space}
The four momenta are defined as $p_+^\mu = \frac{\sqrt{s}}{2}\left(1, \sin \theta, 0, - \cos \theta \right)$,
$p_-^\mu = \frac{\sqrt{s}}{2}\left(1, -\sin \theta, 0, \cos \theta \right)$ with
 $s = (p_+ + p_-)^2$ and $s_{ij} = (k_i + k_j)^2$ while the final state momenta (boosted to a common center of mass
 frame) are
\bea
\frac{2 k_1^\mu}{\sqrt{s_{12}}} &=& \left(\gamma_{12,0} - \gamma_{12} \cos \tilde{\theta}_{12},- \sin \tilde{\theta}_{12} \cos \tilde{\phi}_{12},- \sin \tilde{\theta}_{12} \sin \tilde{\phi}_{12}, \gamma_{12,0} \cos \tilde{\theta}_{12}+ \gamma_{12} \right), \\
\frac{2 k_2^\mu}{\sqrt{s_{12}}} &=&\left(\gamma_{12,0} + \gamma_{12} \cos \tilde{\theta}_{12},\sin \tilde{\theta}_{12} \cos \tilde{\phi}_{12},\sin \tilde{\theta}_{12} \sin \tilde{\phi}_{12}, \gamma_{12,0} \cos \tilde{\theta}_{12}+ \gamma_{12} \right),\\
\frac{2 k_3^\mu}{\sqrt{s_{34}}} &=& \left(\gamma_{34,0} - \gamma_{34} \cos \tilde{\theta}_{34}, \sin \tilde{\theta}_{34} \cos \tilde{\phi}_{34},\sin \tilde{\theta}_{34} \sin \tilde{\phi}_{34}, \gamma_{34,0} \cos \tilde{\theta}_{34}- \gamma_{34} \right),\\
\frac{2 k_4^\mu}{\sqrt{s_{34}}} &=& \left(\gamma_{34,0} + \gamma_{34} \cos \tilde{\theta}_{34}, -\sin \tilde{\theta}_{34} \cos \tilde{\phi}_{34}, -\sin \tilde{\theta}_{34} \sin \tilde{\phi}_{34}, -\gamma_{34,0} \cos \tilde{\theta}_{34}- \gamma_{34} \right).
\eea
We use the definitions $\lambda = s^2 + s_{12}^2 + s_{34}^2 - 2ss_{12}-2ss_{34}-2s_{12}s_{34}$
\begin{align*}
    \gamma_{12} &= \frac{\sqrt{\lambda}}{2\sqrt{s s_{12}}}, & \quad
  \gamma_{12,0} &= \frac{s+s_{12}-s_{34}}{2\sqrt{s s_{12}}}, \\
 \gamma_{34} &= \frac{\sqrt{\lambda}}{2\sqrt{s s_{34}}}, & \quad
  \gamma_{34,0} &= \frac{s+s_{34}-s_{12}}{2\sqrt{s s_{34}}}, \\
  \gamma_{12}^{\pm} &= \gamma_{12,0}\pm\gamma_{12}, & \quad
  \gamma_{34}^{\pm} &= \gamma_{34,0}\pm\gamma_{34}.
\end{align*}
Useful identities are $\gamma_{12,0}^2-\gamma_{12}^2=\gamma_{12}^+\gamma_{12}^-=1$ and
$\gamma_{34,0}^2-\gamma_{34}^2=\gamma_{34}^+\gamma_{34}^- =1$.
A phase convention choice on $\phi_{12,34}$ in the spinors is required to be the
same in the CC03 and CC11 results.

\bibliographystyle{JHEP}
\bibliography{bibliography2}
\end{document}